\documentclass[journal]{IEEEtran}
\usepackage[cmex10]{amsmath}
\usepackage{amssymb,amsthm,amsfonts,mathrsfs}
\usepackage{graphicx}
\usepackage{dsfont,enumitem,stackrel}
\usepackage{color}
\usepackage{mathtools,amssymb,bm}
\usepackage{amstext}
\usepackage{array}
\usepackage{algorithmicx}
\usepackage[ruled]{algorithm}
\usepackage{algpseudocode}
\usepackage{algpascal}
\usepackage{algc}
\usepackage{cases}
\usepackage{pgfplots}
\usepackage{accents}
\usepackage{caption}
\DeclareCaptionLabelSeparator{periodspace}{.\quad}
\usepackage{float}
\usepackage{cite}
\usepackage [autostyle, english = american]{csquotes}
\usepackage{hyperref}
\theoremstyle{remark}

\newcommand\ASTART{\bigskip\noindent\begin{minipage}[b]{0.5\linewidth}}

\newcommand\AENDSKIP{\end{minipage}\bigskip}
\newcommand\AEND{\end{minipage}}
\ifCLASSOPTIONcaptionsoff
  \usepackage[nomarkers]{endfloat}
 \let\MYoriglatexcaption\caption
 \renewcommand{\caption}[2][\relax]{\MYoriglatexcaption[#2]{#2}}
\fi

\theoremstyle{plain}
\newtheorem{thm}{\textbf{Theorem}}
\newtheorem{lem}{\textbf{Lemma}}

\theoremstyle{definition}
\newtheorem{defn}{\textbf{Definition}}

\theoremstyle{remark}
\newtheorem*{rem}{\bf Remark}
\newtheorem*{sketch}{\bf Proof sketch }

\newcommand*{\rom}[1]{\expandafter\@slowromancap\romannumeral #1@}

\graphicspath{{./Figures/}} 
\begin{document}
%
\title{Sample Complexity of Total Variation Minimization}
\author{Sajad~Daei, Farzan~Haddadi, Arash~Amini}%

\maketitle

\begin{abstract}
This work considers the use of Total variation (TV) minimization in the recovery of a given gradient sparse vector from Gaussian linear measurements. It has been shown in recent studies that there exist a sharp phase transition behavior in TV minimization in asymptotic regimes. The phase transition curve specifies the boundary of success and failure of TV minimization for large number of measurements. It is a challenging task to obtain a theoretical bound that reflects this curve. In this work, we present a novel upper-bound that suitably approximates this curve and is asymptotically sharp. Numerical results show that our bound is closer to the empirical TV phase transition curve than the previously known bound obtained by Kabanava.
\end{abstract}


%
\IEEEpeerreviewmaketitle

\section{Introduction}
\IEEEPARstart{C}{ompressed} Sensing (CS) is a method to recover a sparse vector $\bm{x}\in\mathbb{R}^n$ from a few linear measurements
$\bm{y}=\bm{Ax}\in\mathbb{R}^m$ where $\bm{A}\in\mathbb{R}^{m\times n}$ is the measurement matrix. In most cases in practice, the signal $\bm{x}$ is not sparse itself but there exists a dictionary such that $\bm{x}=\bm{D \alpha}$ for some sparse $\bm{\alpha}$. This is known as synthesis sparsity and the following problem called $\ell_1$ minimization in the synthesis form is considered for recovering $\bm{x}$:
\begin{align}\label{l1syn}
\min_{\bm{z}\in\mathbb{R}^n}\|\bm{z}\|_1~\mathrm{s.t.}~\bm{y}=\bm{ADz}.
\end{align}
In \cite{rauhut2008compressed,candes2011compressed,kabanava2015analysis}, recovery guarantees of this problem are studied. In general, one may not be able to correctly estimate $\bm{\alpha}$ from (\ref{l1syn}), but can hope for a good approximation of $\bm{x}=\bm{D \alpha}$ \cite{candes2011compressed}. The second approach to deal with such cases, is to focus on signals that are sparse after the application of an operator called analysis operator $\bm{\Omega}$ (See e.g. \cite{nam2013cosparse,kabanava2015analysis,kabanava2015robust}). In the literature this is known as cosparsity or analysis sparsity. The following problem called $\ell_1$ minimization in the analysis form is studied to estimate the signal $\bm{x}$:
\begin{align}\label{problem.l1analysis}
 &\min_{\bm{z}\in\mathbb{R}^n}\|\bm{\Omega z}\|_1~\mathrm{s.t.}~\bm{y}=\bm{A z}.
 \end{align}
 A special case of this problem that has great importance in a variaty of applications including image processing \footnote{Piecewise constant images are modeled as low variational functions.} is the case where $\bm{\Omega}$ is the one- or two-dimensional difference operator that leads to the total variation (TV) minimization problem which we call $\mathsf{P}_{\mathrm{TV}}$ from this point on.\par
Although many results in the CS literature have been established via Restricted Isometry Property (RIP) and Null Space Property (NSP) conditions (e.g. in \cite{candes2006near,donoho2006high,donoho2011noise,xu2011precise,rauhut2008compressed}), they fail to address gradient sparse\footnote{Low variational signal.} vectors (the rows of the difference matrix do not form a dictionary).\par 
In a separate field of study, it is shown that the problem (\ref{problem.l1analysis}) undergoes a transition from failure to success (known as phase transition) as the number of measurements increases (e.g. see \cite{amelunxen2013living,donoho2013accurate}). Namely, there exist a curve $m=\Psi(s,\bm{\Omega})$ that the problem (\ref{problem.l1analysis}) succeeds to recover a gradient $s$-sparse vector with probability $\frac{1}{2}$. Obtaining a bound that approximates this curve has been an important and challenging task in recent years as it specifies the required number of measurements in problem (\ref{problem.l1analysis}) (See for example \cite{krahmer2017total,donoho2013accurate}). This work revolves around this challenge. Specifically, we propose an upper-bound on $\Psi(s,\bm{\Omega})$ in the case of one dimensional difference operator
\begin{align}
\bm{\Omega}=\begin{bmatrix}
1& -1 & 0 & \cdots & 0 \\
0& 1 & -1 & \cdots & 0  \\
&  \ddots    & \ddots       & \ddots & \\
0 & \cdots & \cdots & 1 & -1 &
\end{bmatrix}\in\mathbb{R}^{n-1\times n}.\nonumber
\end{align}
\subsection{Related Works}
Despite the great importance of TV minimization in imaging sciences, few works have been established to find explicit formula for the number of measurements required for $\mathsf{P}_{\mathrm{TV}}$ to succeed  \cite{krahmer2017total,cai2015guarantees,needell2013stable,donoho2013accurate}. In \cite{needell2013stable}, Needle et al. transformed two-dimensional signals with low variations into those with compressible Haar wavelet coefficients. Then a modified RIP is considered for $\bm{A}$ to guarantee stable recovery. However, their proof does not hold for one-dimensional gradient sparse signals. In \cite{cai2015guarantees}, a geometric approach based on \textquotedblleft escape through a mesh lemma\textquotedblright $~$is used to recover gradient $s$-sparse vectors from Gaussian measurements. Recently, in \cite{krahmer2017total}, Krahmer et al. obtained the number of subgaussian linear measurements in TV minimization based on the mean empirical width \cite{tropp2015convex,vershynin2015estimation}. It is not evident from \cite{krahmer2017total,cai2015guarantees,needell2013stable} whether the obtained lower-bound on the number of measurements is sharp. In \cite{donoho2013accurate}, a lower-bound is derived for TV minimization and its asymptotic sharpness is proved by relating the bound to the normalized Minimum Mean Squared Error (MMSE) of a certain regularized basis pursuit problem (BPDN). In \cite{kabanava2015robust}, an upper-bound on $\Psi(s,\bm{\Omega})$ is proposed. The approach is based on generalizing the proofs of \cite[Proposition 1]{foygel2014corrupted} to TV minimization. 
\subsection{Outline of the paper}
The paper is organized as follows. Section \ref{section.convexgeometry} provides a brief review of some concepts from convex geometry. Section \ref{section.mainresult} discusses our main contribution which determines an upper-bound on the sufficient number of Gaussian measurements for $\mathsf{P}_{\mathrm{TV}}$ to succeed. In Section \ref{section.sim}, numerical experiments are presented to verify our theoretical bound. Finally, the paper is concluded in Section \ref{section.conclusion}.
\subsection{Notation}
Throughout the paper, scalars are denoted by lowercase letters, vectors by lowercase boldface letters, and matrices by uppercase boldface letters. The $i$th element of a vector $\bm{x}$ is shown either by ${x}(i)$ or $x_i$. $(\cdot)^\dagger$ denotes the pseudo inverse operation. We reserve calligraphic uppercase letters for sets (e.g. $\mathcal{S}$). The cardinality of a set $\mathcal{S}$ is shown by $|\mathcal{S}|$. $[n]$ refers to the set $\{1,..., n\}$. Furthermore, we write ${\mathcal{\bar{S}}}$ for the complement $[n]\setminus\mathcal{
	S}$ of a set $\mathcal{S}$ in $[n]$. For a matrix $\bm{X}\in\mathbb{R}^{m\times n}$ and a subset $\mathcal{S}\subseteq [n]$, the notation $\bm{X}_\mathcal{S}$ is used to indicate the row submatrix of $\bm{X}$ consisting of the rows indexed by $\mathcal{S}$. Similarly, for $\bm{x}\in\mathbb{R}^n$, $\bm{x}_\mathcal{S}$ is the subvector in $\mathbb{R}^{|\mathcal{S}|}$ consisting of the entries
indexed by $\mathcal{S}$, that is, $(\bm{x}_S)_i = x_{j_i}~:~\mathcal{S}=\{j_i\}_{i=1}^{|\mathcal{S}|}$. Lastly, the polar $\mathcal{K}^{\circ}$ of a cone $\mathcal{K}\subset\mathbb{R}^n$ is the set of vectors forming non-acute angles with every vector in $\mathcal{K}$, i.e. \begin{align}
\mathcal{K}^\circ=\{\bm{v}\in\mathbb{R}^n: \langle \bm{v}, \bm{z} \rangle\le 0~\forall \bm{z}\in\mathcal{K}\}.
\end{align}
\section{Convex Geometry}\label{section.convexgeometry}
In this section, basic concepts of convex geometry are reviewed.
\subsection{Descent Cones}
The descent cone of a proper convex function $f:\mathbb{R}^n\rightarrow \mathbb{R}\cup \{\pm\infty\}$ at point $\bm{x}\in \mathbb{R}^n$ is the set of directions from $\bm{x}$ in which $f$ does not increase:
\begin{align}\label{eq.descent cone}
\mathcal{D}(f,\bm{x})=\bigcup_{t\ge0}\{\bm{z}\in\mathbb{R}^n: f(\bm{x}+t\bm{z})\le f(\bm{x})\}\cdot
\end{align}
The descent cone of a convex function is a convex set. There is a famous duality result \cite[Ch. 23]{rockafellar2015convex} between the decent cone and the subdifferential of a convex function  given by:
\begin{align}\label{eq.D(f,x)}
\mathcal{D}^{\circ}(f,\bm{x})=\mathrm{cone}(\partial f(\bm{x})):=\bigcup_{t\ge0}t.\partial f(\bm{x}).
\end{align}
\subsection{Statistical Dimension}
\begin{defn}{(Statistical Dimension\cite{amelunxen2013living})}. Let $\mathcal{C}\subseteq\mathbb{R}^n$ be a convex closed cone. The statistical dimension of $\mathcal{C}$ is defined as:
	\begin{align}\label{eq.statisticaldimension}
	\delta(\mathcal{C}):=\mathds{E}\|\mathcal{P}_\mathcal{C}(\bm{g})\|_2^2=\mathds{E}\mathrm{dist}^2(\bm{g},\mathcal{C}^\circ),
	\end{align}
	where, $\mathcal{P}_\mathcal{C}(\bm{x})$ is the projection of $\bm{x}\in \mathbb{R}^n$ onto the set $\mathcal{C}$ defined by: $\mathcal{P}_\mathcal{C}(\bm{x})=\underset{\bm{z} \in \mathcal{C}}{\arg\min}\|\bm{z}-\bm{x}\|_2$.
\end{defn}
The statistical dimension generalizes the concept of dimension for subspaces to the class of convex cones. Let $f$ be a function that promotes some low-dimensional structure of $\bm{x}$. Then, $\delta(\mathcal{D}(f,\bm{x}))$ specifies the required number of Gaussian measurements that the optimization problem 
\begin{align}
&\min_{\bm{z}\in\mathbb{R}^n}f(\bm{z})\nonumber\\
&\mathrm{s.t.}~~ \bm{y}=\bm{A}\bm{z},
\end{align}
needs for successful recovery \cite[Theorem 2]{amelunxen2013living}.
\section{Main result}\label{section.mainresult}
In this work, we provide an upper-bound for the required number of Gaussian measurements for $\mathsf{P}_{\mathrm{TV}}$ to succeed. The result is summarized in the following theorem.
\begin{thm}\label{theorem.main}
Let $\bm{x}\in\mathbb{R}^n$ be a gradient $s$-sparse vector with gradient support $\mathcal{S}$. Let $\bm{A}$ be an $m\times n$ matrix whose rows are independent random vectors drawn from $\mathcal{N}(\bm{0},\bm{I}_n)$, and let $\bm{y}=\bm{A x}+\bm{e}\in\mathbb{R}^m$ be the vector of measurements. Assume that $\|\bm{e}\|_2\le \eta$ and let $\hat{\bm{x}}_{\eta}$ be any solution of $\mathsf{P}_{\mathrm{TV}}$. Then,
\begin{align}\label{eq.upperboundstatic}
\inf_{t\ge 0}\mathds{E}\mathrm{dist}^2(\bm{g},t\partial\|\cdot\|_{\mathrm{TV}}(\bm{x}))\le n-\frac{3(n-1-s)^2}{\pi(2n+s-4)},
\end{align}
and if 
\begin{align}\label{eq.mainbound}
m>\Bigg(\sqrt{n-\frac{3(n-1-s)^2}{\pi(2n+s-4)}}+t+\tau\Bigg)^2+1,
\end{align}
then, the following statement holds:
\begin{align}
\|\hat{\bm{x}}_{\eta}-\bm{x}\|_2\le \frac{2\eta}{\tau},
\end{align}
with probability at least $1-e^{-\frac{t^2}{2}}$.
\end{thm}
\begin{sketch}
The left-hand side of (\ref{eq.upperboundstatic}), besides the infimum over $t$, implicitly includes an infimum over the set $\partial\|\cdot\|_{\mathrm{TV}}(\bm{x})$ because of the definition of "dist". Instead of this latter infimum, we choose a vector in the set $\partial\|\cdot\|_{\mathrm{TV}}(\bm{x})$ that leads to an upper bound for $\mathrm{dist}^2(\bm{g},t\partial\|\cdot\|_{\mathrm{TV}}(\bm{x}))$. This results in a strictly convex function of $t$. Then, by finding infimum over $t$, we obtain the desired upper-bound.
\end{sketch}
See Appendix \ref{proof.theoremmain} for details.
\begin{rem}
In \cite[Lemma 1]{kabanava2015robust}, the following upper-bound is derived for $\delta(\mathcal{D}(\|\cdot\|_{\mathrm{TV}},\bm{x}))$:
\begin{align}\label{eq.kab}
\inf_{t\ge 0}\mathds{E}\mathrm{dist}^2(\bm{g},t\partial\|\cdot\|_{\mathrm{TV}}(\bm{x}))\le
n-\frac{(n-1-s)^2}{n\pi}.
\end{align}
This bound is rather loose in low sparsity regimes. The main ideas in the proof of this bound are drawn from \cite[Proposition 1]{foygel2014corrupted}.
\end{rem}
\subsection{Discussion}
In \cite[Proposition 1]{foygel2014corrupted}, an upper-bound is derived for $\delta(\mathcal{D}(f,\bm{x}))$ where $f$ is a decomposable norm function\footnote{See \cite[Section 2.1]{candes2013simple} for more explanations.} that promotes a low-dimensional structure. This upper-bound does not approach the phase transition curve in the low-dimensional structured regimes. The problem arises from a redundant maximization in the proof that increases the number of required measurements (See section \ref{section.sim}). In Theorem \ref{theorem.main}, we propose a tighter upper-bound that leads to a reduction in the required number of Gaussian measurements in TV minimization. This upper-bound better follows the empirical TV phase transition curve. The upper-bound and the proof approach are completely new and differ from \cite{kabanava2015robust} and \cite{foygel2014corrupted}. Our bound only depends on the sparsity level $s$ and the special properties of the difference operator $\bm{\Omega}$. It also tends to the empirical TV phase transition curve at large values of $m$. In addition to TV, our approach can be applied to other low dimensional structures. For instance, the result in Theorem \ref{theorem.main} can be easily extended to two dimensional images. Compared with \cite[Theorem 5]{kabanava2015robust}, the reduction of the required number of measurements, would be more evident in that case.
\section{Numerical Experiments}\label{section.sim}
\begin{figure}[t]
	\hspace*{-.5cm}
	\includegraphics[scale=.28]{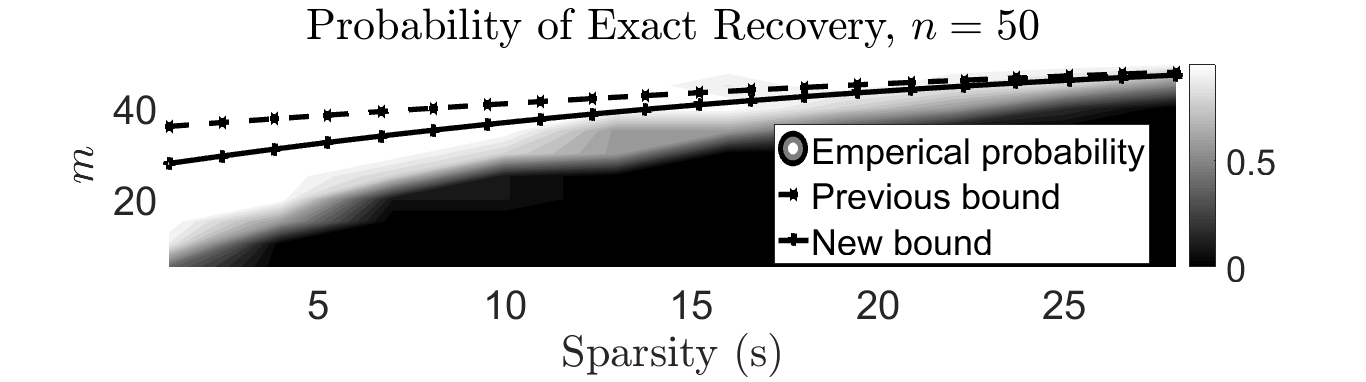}
	\caption{Phase transition of $\mathsf{P}_{\mathrm{TV}}$ in the case of $n=50$. The empirical probability is computed over $50$ trials (black=$0 \%$, white=$100\%$). The previous and new bounds come from (\ref{eq.kab}) and (\ref{eq.upperboundstatic}), respectively. } 
	\label{fig.tvphase3}
\end{figure}

\begin{figure}[t]
	\hspace*{-.5cm}
	\includegraphics[scale=.28]{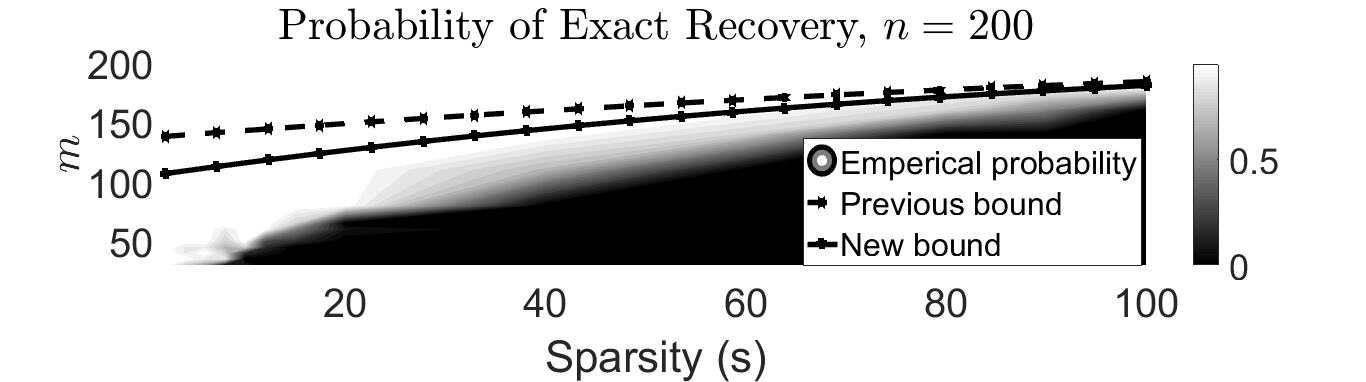}
	\caption{Phase transition of $\mathsf{P}_{\mathrm{TV}}$ in the case of $n=200$. The empirical probability is computed over $50$ trials (black=$0 \%$, white=$100\%$). The previous and new bounds come from (\ref{eq.kab}) and (\ref{eq.upperboundstatic}), respectively. }
	\label{fig.tvphase1}
\end{figure}

\begin{figure}[t]
	\hspace*{-.5cm}
	\includegraphics[scale=.28]{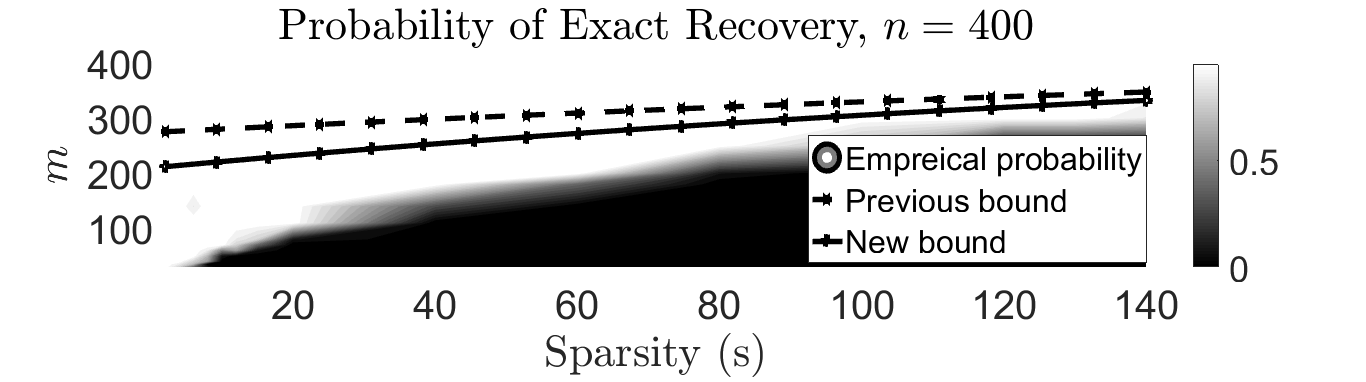}
	\caption{Phase transition of $\mathsf{P}_{\mathrm{TV}}$ in the case of $n=500$. The empirical probability is computed over $50$ trials (black=$0 \%$, white=$100\%$). The previous and new bounds come from (\ref{eq.kab}) and (\ref{eq.upperboundstatic}), respectively. }
	\label{fig.tvphase2}
\end{figure}

In this section, we evaluate how the number of Gaussian measurements scales with gradient sparsity. For each $m$ and $s$, we repeat the following procedure $50$ times in the cases $n=50$, $n=200$ and $n=400$:
\begin{itemize}
	\item Generate a vector $\bm{x}\in\mathbb{R}^n$ that its discrete gradient has $s$ non-zero entries. The locations of the non-zeros are selected at random.
	\item Observe the vector $\bm{y}=\bm{A}\bm{x}$, where $\bm{A}\in\mathbb{R}^{m\times n}$ is a random matrix whose elements are drawn from an i.i.d standard Gaussian distribution.
	\item Obtain an estimate $\hat{\bm{x}}$ by solving $\mathsf{P}_{\mathrm{TV}}$.
	\item Declare success if $\|\bm{x}-\hat{\bm{x}}\|_2\le 10^{-3}$.
\end{itemize}
Figs. \ref{fig.tvphase3} \ref{fig.tvphase1} and \ref{fig.tvphase2} show the empirical probability of success for this procedure. As shown in Figs. \ref{fig.tvphase3} \ref{fig.tvphase1} and \ref{fig.tvphase2}, our new bound better describes $\delta(\mathcal{\|\cdot\|_{\mathrm{TV}}},\bm{x})$ in particular in low sparsity regimes. As sparsity increases, the difference between our bound and the bound (\ref{eq.kab}) gets less. When the dimension of the true signal i.e. $n$, increases, the difference between our bound and (\ref{eq.kab}) enhances (See Figs. \ref{fig.tvphase3}, \ref{fig.tvphase1} and \ref{fig.tvphase2}). In the asymptotic case, it seems that our bound reaches the empirical TV phase transition curve.
\section{Conclusion}\label{section.conclusion}
We have investigated the nonuniform recovery of gradient sparse signals from Gaussian random measurements. Obtaining a bound that suitably describes the precise behavior of TV minimization from failure to success, is left as an unanswered question. In this work, we derived an upper-bound for the required number of measurements that approximately estimates this behavior. Also, this bound is close to the empirical TV phase transition curve and seems to be asymptotically sharp. 
\appendices
\section{Proofs}
\subsection{Proof of Theorem \ref{theorem.main}}\label{proof.theoremmain}
\begin{proof}
Fix $\bm{g}\in\mathbb{R}^n$. Define
\begin{align}
&s_1=\{\#i\in\{2,...,n-1\}:~i\in \mathcal{S}, i-1\in \mathcal{S}\},\nonumber\\
&s_2=\{\#i\in\{2,...,n-1\}:~i\in \mathcal{\bar{S}}, i-1\in \mathcal{\bar{S}}\}.
\end{align}
Since $\partial \|\cdot\|_1(\bm{\Omega x})$ is a compact set, for any $\bm{z}\in\partial \|\cdot\|_1(\bm{\Omega x})$, there exists a $\bm{z}_0\in\partial \|\cdot\|_1(\bm{\Omega x})$ such that:
\begin{align}
\bm{z}_0=\arg\max_{\bm{z}\in\partial\|\cdot\|_1(\bm{\Omega x})} \langle \bm{g}, \bm{\Omega}^T\bm{z} \rangle=\mathrm{sgn}(\bm{\Omega x})_{\mathcal{S}}+\mathrm{sgn}(\bm{\Omega g})_{\mathcal{\bar{S}}}.
\end{align} 
Then, we have:
\begin{align}
&\mathrm{dist}^2(\bm{g},t\bm{\Omega}^T\partial\|\cdot\|_1(\bm{\Omega x}))\le\|\bm{g}-t\bm{\Omega}^T\bm{z}_0\|_2^2=\nonumber\\
&\|\bm{g}-t\bm{\Omega}_{\mathcal{S}}^T\mathrm{sgn}(\bm{\Omega}\bm{x})_{\mathcal{S}}-t\bm{\Omega}_{\mathcal{\bar{S}}}^T\mathrm{sgn}(\bm{\Omega}\bm{g})_{\mathcal{\bar{S}}}\|_2^2=\|\bm{g}\|_2^2+\nonumber\\
&t^2\|\bm{\Omega}_{\mathcal{S}}^T\mathrm{sgn}(\bm{\Omega}\bm{x})_{\mathcal{S}}\|_2^2+t^2\|\bm{\Omega}_{\mathcal{\bar{S}}}^T\mathrm{sgn}(\bm{\Omega}\bm{g})_{\mathcal{\bar{S}}}\|_2^2\nonumber\\
&-2t\langle \bm{g},\bm{\Omega}_{\mathcal{\bar{S}}}^T\mathrm{sgn}(\bm{\Omega}\bm{g})_{\mathcal{\bar{S}}} \rangle+2t^2\langle \bm{\Omega}_{\mathcal{S}}^T\mathrm{sgn}(\bm{\Omega}\bm{x})_{\mathcal{S}},\bm{\Omega}_{\mathcal{\bar{S}}}^T\mathrm{sgn}(\bm{\Omega}\bm{g})_{\mathcal{\bar{S}}}\rangle.
\end{align}
By taking expectation from both sides, we have:
\begin{align}\label{eq.upperbound}
&\mathds{E}\|\bm{g}-t\bm{\Omega}_{\mathcal{S}}^T\mathrm{sgn}(\bm{\Omega}\bm{x})_{\mathcal{S}}-t\bm{\Omega}_{\mathcal{\bar{S}}}^T\mathrm{sgn}(\bm{\Omega}\bm{g})_{\mathcal{\bar{S}}}\|_2^2=\nonumber\\
&n-2t\sqrt{\frac{2}{\pi}}\sum_{i\in\mathcal{\bar{S}}}\|\bm{\omega}_i\|_2+t^2\Big[\sum_{j\in \mathcal{S}}\sum_{k\in\mathcal{S}}\bm{\omega}_j^T\bm{\omega}_k\mathrm{sgn}(\bm{\Omega x})_j\mathrm{sgn}(\bm{\Omega x})_k\Big]\nonumber\\
&+t^2\mathds{E}\Big[\sum_{j\in \mathcal{\bar{S}}}\sum_{k\in\mathcal{\bar{S}}}\bm{\omega}_j^T\bm{\omega}_k\mathrm{sgn}(\bm{\Omega g})_j\mathrm{sgn}(\bm{\Omega g})_k\Big]\stackrel{(1)}{\le}\nonumber\\
&n-2t\sqrt{\frac{2}{\pi}}\sum_{i\in\mathcal{\bar{S}}}\|\bm{\omega}_i\|_2+t^2\Big[\sum_{j\in \mathcal{S}}\sum_{k\in\mathcal{S}}\bm{\omega}_j^T\bm{\omega}_k\mathrm{sgn}(\bm{\Omega x})_j\mathrm{sgn}(\bm{\Omega x})_k\Big]\nonumber\\
&+t^2\Big[\sum_{j\in \mathcal{\bar{S}}}\sum_{k\in\mathcal{\bar{S}}}\bm{\omega}_j^T\bm{\omega}_k\frac{2}{\pi}\mathrm{sin}^{-1}\frac{\bm{\omega}_j^T\bm{\omega}_k}{\|\bm{\omega}_j\|_2\|\bm{\omega}_k\|_2}\Big]\stackrel{(2)}{\le}\nonumber\\
&n-\frac{4t}{\sqrt{\pi}}\bar{s}+t^2[2s+2s_1+2\bar{s}+\frac{2s_2}{3}]\stackrel{(3)}{\le}\nonumber\\
&n-\frac{4t}{\sqrt{\pi}}\bar{s}+t^2[4s-2+\frac{8}{3}\bar{s}-\frac{2}{3}],
\end{align}	
where $(1)$ follows from the following lemma and $\bm{\Omega}:=[\bm{\omega}_1, \bm{\omega}_2,..., \bm{\omega}_p]^T$.
\begin{lem}\label{lemma.Esign}
	Let $\bm{g}\in\mathbb{R}^n$ be a standard random Gaussian i.i.d vector and $\bm{\Omega}\in\mathbb{R}^{p\times n}$ be an analysis operator. Then,
	\begin{align}
	\mathds{E}\{\mathrm{sgn}(\bm{\Omega g})_j\mathrm{sgn}(\bm{\Omega g})_k\}=\frac{2}{\pi}\mathrm{sin}^{-1}\frac{\bm{\omega}_j^T\bm{\omega_k}}{\|\bm{\omega}_j\|_2\|\bm{\omega}_k\|_2}.
	\end{align}	
\end{lem}
Proof. see Appendix \ref{proof.Esign}.\\
The inequality $(2)$ is the result of the following properties of the difference operator. 
\begin{align}
&\bm{\omega}_j^T\bm{\omega}_k = \left\{\begin{array}{lr}
-1, &  |j-k|=1\\
0, & \mathrm{o.w.}
\end{array}\right\},\nonumber\\
&\|\bm{\omega}_i\|_2=\sqrt{2}~:~\forall i\in {1,..., n-1}.\nonumber\\
\end{align}
Also, $\bar{s}=n-1-s$. The inequality $(3)$ comes from the facts
\begin{align}
&s_1\le s-1,\nonumber\\
&s_2\le \bar{s}-1.\nonumber\\
\end{align}
Now, by minimizing (\ref{eq.upperbound}) with respect to $t$, we reach (\ref{eq.upperboundstatic}).
Due to \cite[Corolarry 3.5]{tropp2015convex}, if 
\begin{align}\label{eq.mohem}
m>(\sqrt{\delta(\mathcal{D}(\|\cdot\|_{\mathrm{TV}},\bm{x}))}+t+\tau)^2+1,
\end{align}
then, with probability $1-e^{-\frac{t^2}{2}}$,
\begin{align}
\|\bm{x}-\hat{\bm{x}}\|_2\le \frac{2\eta}{\tau}
\end{align}
A good upper-bound for $\delta(\mathcal{D}(\|\cdot\|_{\mathrm{TV}},\bm{x}))$, is given by (\ref{eq.upperboundstatic}) and thus, the claim is proved.
\end{proof}
\subsection{Proof of Lemma \ref{lemma.Esign}}\label{proof.Esign}
\begin{proof}
	Consider $\bm{\Omega}:=[\bm{\omega}_1, \bm{\omega}_2,..., \bm{\omega}_p]^T$.
	Define
	\begin{align}
	&{h}_j=\frac{\bm{\omega}_j^T\bm{g}}{\|\bm{\omega}_j\|_2},\nonumber\\
	&{h}_k=\frac{\bm{\omega}_k^T\bm{g}}{\|\bm{\omega}_k\|_2}.
	\end{align}	
	We have:
	\begin{align}
	&\mathds{E}\{\mathrm{sgn}(\bm{\Omega g})_j\mathrm{sgn}(\bm{\Omega g})_k\}=\mathds{E}\{\mathrm{sgn}({h}_j)\mathrm{sgn}({h}_k)\}=\nonumber\\
	&1-2\mathds{P}\{\frac{h_j}{h_k}<0\}=1-2(\frac{1}{2}-\frac{1}{\pi}\mathrm{sin}^{-1}(\frac{\mathds{E}\{\bm{\omega}_k^T\bm{g}\bm{\omega}_j^T\bm{g}\}}{\|\bm{\omega}_j\|_2\|\bm{\omega}_k\|_2}))=\nonumber\\
	&\frac{2}{\pi}\mathrm{sin}^{-1}\frac{\bm{\omega}_j^T\bm{\omega_k}}{\|\bm{\omega}_j\|_2\|\bm{\omega}_k\|_2}
	\end{align}
	where the second equality comes from total probability theorem, the third equality comes from the fact that $\frac{h_j}{h_k}$ is a Cauchy random variable.
\end{proof}
\ifCLASSOPTIONcaptionsoff
  \newpage
\fi

\bibliographystyle{ieeetr}
\bibliography{mypaperbibe}
\end{document}